\documentclass[a4paper]{article}

\usepackage[english]{babel}
\usepackage[utf8]{inputenc}
\usepackage{amsmath}
\usepackage{graphicx}
\usepackage[colorinlistoftodos]{todonotes}
\usepackage[a4paper]{geometry}
\usepackage{braket}
\usepackage[hidelinks]{hyperref}
\usepackage{hyperref}
\usepackage{comment}
 
\title{Reformulating Bell's Theorem: The Search for a Truly Local Quantum Theory\thanks{This is a preprint. The final and definitive version of this paper will be published in \textit{Studies in History and Philosophy of Modern Physics}.}} 
 
\author{Mordecai Waegell\thanks{Institute for Quantum Studies, Chapman University, CA 92866, United States} \space and Kelvin J. McQueen\thanks{Department of Philosophy, Chapman University, CA 92866, United States.}}

\date{\today}

\begin{document}
\maketitle

\begin{abstract}
The apparent nonlocality of quantum theory has been a persistent concern. Einstein et. al. (1935) and Bell (1964) emphasized the apparent nonlocality arising from entanglement correlations. While some interpretations embrace this nonlocality, modern variations of the Everett-inspired many worlds interpretation try to circumvent it. In this paper, we review Bell's ``no-go" theorem and explain how it rests on three axioms, \textit{local causality}, \textit{no superdeterminism}, and \textit{one world}. Although Bell is often taken to have shown that \textit{local causality} is ruled out by the experimentally confirmed entanglement correlations, we make clear that it is the conjunction of the three axioms that is ruled out by these correlations. We then show that by assuming \textit{local causality} and \textit{no superdeterminism}, we can give a direct proof of many worlds. The remainder of the paper searches for a consistent, local, formulation of many worlds. We show that prominent formulations whose ontology is given by the wave function violate \textit{local causality}, and we critically evaluate claims in the literature to the contrary.  We ultimately identify a local many worlds interpretation that replaces the wave function with a separable Lorentz-invariant wave-field.  We conclude with discussions of the Born rule, and other interpretations of quantum mechanics.

\end{abstract}

\tableofcontents

\section{Introduction}

In the early days of quantum theory, many physicists and philosophers were troubled by the apparent nonlocality of the theory.  The first major objection (expressed by Einstein at the 1927 Solvay Conference) was that if the wavefunction is interpreted as an extended physical object in space-time, then the proposed process of wavefunction collapse must occur instantaneously everywhere that the wavefunction exists---a manifestly nonlocal process, which seems at odds with special relativity. Many interpretations of quantum mechanics have arisen since those early days, and some have embraced this type of nonlocality, such as the dynamical collapse theories.\footnote{See e.g. Pearle (1976) and Ghirardi, Rimini, and Weber (1986).} Many other models circumvent it by removing wavefunction collapse from the theory. 

Quantum entanglement produces another type of nonlocality, as discussed by Einstein et. al. (1935) and Bell (1964). This nonlocality manifests when the empirical correlations between remote systems cannot be explained by local common causes, which again seems at odds with special relativity. Some interpretations of quantum mechanics embrace this nonlocality, such as Bohmian mechanics (Bohm 1952) and the dynamical collapse theories, while others attempt to circumvent it. The most prominent such attempts are the various many worlds interpretations inspired by Everett (1957). These attempts to circumvent the nonlocality of entanglement correlations are the primary focus of this paper.

In Sec. \ref{Locality} we review the EPR argument of Einstein et. al. and introduce a number of formal definitions regarding locality and elements of reality, which are then used throughout the paper. The most crucial notion is that of \textit{local causality}, which states that there can be no cause and effect between space-like separated events. A theory which violates this axiom is nonlocal.

In Sec. \ref{Bell} we review Bell's ``no-go" theorem and (following others, e.g. Myrvold et. al. (2019 sec. 3.2)) we explain how the theorem concerns \textit{local causality} and two other implicit assumptions. We make Bell's assumptions explicit and refer to them as the \textit{no superdeterminism} axiom and the \textit{one-world} axiom. The former axiom effectively rules out explanations of the entanglement correlations that appeal to either effects preceding causes or fine-tuning. The latter axiom is effectively a denial of many worlds. Making these axioms explicit enables one to reformulate Bell's theorem. The new formulation is more of a \textit{which-way} theorem than a \textit{no-go} theorem, since the thesis is that at least one of its three axioms must be violated by any physical theory that reproduces the empirical entanglement correlations. With these axioms made explicit, Bell's original proof of his no-go theorem can be interpreted as proving that a model which obeys \textit{one world} and \textit{no superdeterminism} must violate \textit{local causality}. It logically follows that if \textit{any} two of these axioms is true, then the third must be false. Thus, assuming \textit{local causality} and \textit{no superdeterminism} provides an indirect proof of many worlds (Myrvold et. al. (2019, sec. 8.1.1.)).

In Sec. \ref{BellMany} we provide a direct proof of many worlds from \textit{local causality}, \textit{no superdeterminism} and the relevant empirical facts (i.e. the observed outcomes of measurements in a laboratory). The direct proof explicitly shows that after a quantum experiment there must be multiple copies of each experimenter. This is philosophically significant since it seems to provide an explanation of many worlds, which may make it more accessible to a skeptical audience. As with Bell's proof, our proof makes no use of the quantum formalism -- only the axioms of Bell's theorem and the relevant empirical facts.

Since we are interested in models that obey both \textit{local causality} and \textit{no superdeterminism}, it will be useful for brevity to introduce the term \textit{truly local} to refer to models that satisfy both these axioms. In \textit{truly local} models all causes are within the backwards light-cone of their effects. In the next three sections (5-7) we examine a number of alternative many worlds models, and show that almost none of them are \textit{truly local}, meaning they must also violate either \textit{local causality} or \textit{no superdeterminism}. Sections 5 and 6 deal with prominent $\psi$-ontic models whose ontology is (or is represented by) the universal wavefunction. Section 7 then deals with a model that replaces the wavefunction with a separable Lorentz-invariant wave-field. We will argue that it is only by replacing the wavefunction that we can obtain a \textit{truly local} model.

In Sec. \ref{Global} we consider models that postulate many \textit{global worlds}. A global world is one in which every macroscopic property has a single definite value. We begin with branching models where branching happens everywhere if it happens somewhere (Sec. \ref{Global branching}), creating many global worlds. We then consider non-branching (``divergent") models where all global worlds have always existed (Sec. \ref{sec:GlobDiv}). By reformulating Bell's theorem once again, we prove that none of these models are \textit{truly local}.

In Sec. \ref{Semi} we consider models that postulate \textit{semi-local worlds}. Semi-local worlds are worlds created by branching, where the branching only occurs locally for the systems directly involved in the measurement, and then spreads out through the background space-time as a wave-front whose speed is bounded by $c$. In Sec. \ref{sec:SLocBranch} we show that these models cannot be \textit{truly local} because they assume a single universal wavefunction. In Sec. \ref{Oxford}, we consider the arguments of the Oxford Everettians, who adopt a semi-local model but argue that it is local. We critically examine these arguments.

In Sec. \ref{LocalWorlds} we consider models that postulate \textit{local worlds}. Local models exhibit the same local branching structure as semi-local models, but replace the universal wavefunction in configuration space with a separable Lorentz-invariant wave-field in space-time. This provides an explicit example of a \textit{truly local} many worlds model.  Models of this type have yet to be extensively developed, and are mostly referred to as \textit{parallel lives} interpretations in the literature.

In Sec. \ref{Discussion} we expand the discussion in several directions. In Sec. \ref{Born} we discuss proofs of the Born rule in global and semi-local models that rely on locality principles, and we explain why local models do not need to prove the Born rule. Finally, in Sec. \ref{Interps} we review a number of modern interpretations of quantum mechanics. All such interpretations may be categorized in terms of which axiom(s) they deny --- and the only way to evade this fact would be to find an explicit loophole or logical flaw in the proof of the theorem. We also address some interpretations that seem to deny or ignore the fact that \textit{truly local} models must have many worlds. We conclude with some final remarks in Sec. \ref{sec:Conc}.

\section{Local causality} \label{Locality}

In their celebrated 1935 paper, Einstein, Podolsky, and Rosen [EPR] challenged the \textit{completeness} of quantum mechanics. For a description of the world to be complete, on their definition, means that no \textit{element of reality} is left out of that description. Thus, they sought to show that some elements of reality are left out of the quantum mechanical description of the world.

EPR did not define \textit{element of reality}. Instead, their argument depends on a proposed \textit{sufficient condition} on what it takes for something to be an element of reality. (The only other assumption they make in their argument is that the predictions of quantum mechanics are all correct.) We will refer to the proposed sufficient condition as the \textit{$\lambda$-criterion}. 

\begin{quote}
\textbf{$\lambda$-criterion}: if, without in any way disturbing a system, we can predict with certainty (i.e. with probability equal to unity) the value of a physical quantity possessed by that system, then there exists an element of physical reality ($\lambda$) corresponding to this physical quantity.
\end{quote}

EPR therefore sought to demonstrate circumstances where quantum mechanics predicts with certainty the value of a system's physical quantity, despite failing to describe the underlying element of reality. Here is an example of such a circumstance. Consider particles \textit{a} and \textit{b}, which are separated by some distance. Their spins are entangled as follows:

\begin{equation}\label{EPR}
\Ket{\textrm{EPR}} = \frac{1}{\sqrt 2}\Big( \Ket{\uparrow_z}_{\rm a}\Ket{\uparrow_z}_{\rm b}-\Ket{\downarrow_z}_{\rm a}\Ket{\downarrow_z}_{\rm b}\Big)
\end{equation}

Alice measures the \textit{z}-spin of particle \textit{a}. The two possible outcomes have equal probability of occuring. We now imagine that Bob is about to measure the \textit{z}-spin of particle \textit{b}. Quantum mechanics predicts with certainty that Bob will get the same result as Alice. This is an \textit{entanglement correlation} of quantum mechanics.

EPR now make a crucial \textit{locality assumption}: Alice's measurement cannot disturb Bob's particle because Bob and his particle are (arbitrarily) far away. But the instant that Alice measures her particle, she can use the entanglement correlation to predict with certainty the value of Bob's $z$-spin measurement before he performs it. So by the $\lambda$-criterion, there must already exist an element of reality ($\lambda_Z^B$) corresponding to the value of Bob's $z$-spin measurement. But then equation (\ref{EPR}) is an incomplete description: it fails to define $\lambda_Z^B$. That's the argument.

Note that we can write equation (\ref{EPR}) down in a different basis and run the exact same argument. For example, 

\begin{equation}\label{EPR2}
\Ket{\textrm{EPR}} = \frac{1}{\sqrt 2}\Big( \Ket{\uparrow_y}_{\rm a}\Ket{\uparrow_y}_{\rm b}-\Ket{\downarrow_y}_{\rm a}\Ket{\downarrow_y}_{\rm b}\Big)
\end{equation}

In fact, we can run the same argument for any of the continuous infinity of incompatible electron spin observables. We may then conclude that equation (\ref{EPR}) is infinitely incomplete, since no matter what basis we write it down in, it fails to describe an infinity of elements of reality.

EPR seemingly discovered that the entanglement correlations of quantum mechanics are inconsistent with the locality assumption that is implicit in the $\lambda$-criterion. For the entanglement correlations suggest that when Alice measures the $z$-spin of her particle, then no matter how far away Bob is, the $z$-spin of Bob's particle is \textit{instantaneously} determined. EPR could not believe that nature involved such nonlocal ``action at a distance". 

EPR did not spell out the idea of local causality in any detail. In what follows, we will work with one natural way of spelling it out (in Sec. \ref{Oxford} we consider some related notions of locality). A natural explication is:

\begin{quote}
\textbf{Local causality}: There can be no cause and effect between space-like separated events.
\end{quote}

An \textit{event} is a single point in spacetime. Two events are \textit{space-like separated} if there exists a Lorentz frame where the two events occur simultaneously, but in different places. \textit{Local causality} then captures the idea that Alice cannot instantaneously ``disturb" Bob's system given that Alice and Bob's system are (arbitrarily) far apart.  In general, it means that no intervention at one event can produce a change of any kind at another space-like separated event.

There is also a notion of locality implicit in EPR's definition of an element of reality. To make this explicit, we begin with a more general definition that is independent of \textit{local causality}:

\begin{quote}
\textbf{Element of reality}: If a system's response to an intervention can be predicted with certainty, then there is an element of reality which determines that response.
\end{quote}

An \textit{element of reality} can exist for general interventions and responses and the definition is completely independent of cause and effect in space-time -- it is just a name we assign to a fact of nature.  To capture the EPR notion, we define a subclass of \textit{elements of reality} which explicitly obeys \textit{local causality}:

\begin{quote}
\textbf{Localized element of reality}: If an intervention and response happen in a finite region of space-time, and the response can be predicted with certainty, then there is an element of reality located only in that region that determines that response.
\end{quote}

The \textit{localized element of reality} must be confined to this finite region; if it were not, then the response could be affected by causes in a space-like separated region, violating \textit{local causality}. With this restricted definition, the element which determines Bob's outcome must be located entirely within Bob's region, which is the other piece of the EPR definition.

\section{Reformulating Bell's theorem} \label{Bell}

In 1964, John Stuart Bell realized that the existence of EPR's elements of reality make an experimental difference. In fact, he found that their existence - together with some additional (apparently innocuous) assumptions - contradicts the predictions of quantum mechanics. We axiomatize these assumptions as follows:

\begin{quote}

\textbf{Local Causality}: There can be no cause and effect between space-like separated events.

\textbf{One World}: There exists a single-world ontology in which every measurement has a single definite outcome.

\textbf{No Superdeterminism}: The elements of reality that determine a system's response to interventions do not determine what interventions will occur on that system. 

\end{quote}

In Bell's original theorem, \textit{local causality} was stated as a hypothesis (1964, p195), \textit{no superdeterminism} was implicit with the exception of an afterthought in the conclusion (1964, p199), and \textit{one world} was an unmentioned implicit assumption. The thesis of the reformulated theorem is that at least one of these three axioms must be violated by any physical theory which reproduces the empirical entanglement correlations.

The \textit{one world} axiom is a straightforward denial of many worlds.  The \textit{no superdeterminism} axiom entails that the set of future interventions on a system is independent of that system's present elements of reality. More generally, it enforces the principle that things that should not be correlated are not correlated, e.g. the results of two classical coin flips. This axiom forbids such conspiratorial correlations. 

Since \textit{one world} and \textit{no superdeterminism} are often taken as sacrosanct, it is the third axiom, \textit{local causality}, which is usually given up, leading to the statement that quantum mechanics is a nonlocal theory. For example, the moral drawn by Albert (1992, ch.3):

\begin{quote}
``What Bell has given us is a proof that there is as a matter of fact a genuine nonlocality in the actual workings of nature, \textit{however} we attempt to describe it, period. That nonlocality is, to begin with, a feature
of quantum mechanics itself, and it turns out (via Bell's theorem)
that it is necessarily also a feature of every possible manner of
calculating (without or with superpositions) which produces the
same statistical predictions as quantum mechanics does; and those
predictions are now experimentally known to be correct [...] and so the assumption
that the physical workings of the world are invariably local must
(astonishingly) be false." 
\end{quote}

Here, we prove the which-way theorem by showing that obeying \textit{true locality} and the empirical entanglement correlations requires a many-worlds ontology, and thus violation of the \textit{one world} axiom. Our proof makes use of several preliminary concepts.  First, we say there is a \textit{reproducible perfect correlation} between two systems if and only if the observed outcome of a measurement on one system is necessary for the observed outcome of a measurement on the other system to be predicted with certainty, every time the systems are prepared and measured in the same way.

We assume that a reproducible perfect correlation between two systems must have a cause, and so should not be attributed to coincidence.  Then, if such a correlation exists between two space-like separated events,  \textit{local causality} requires that the cause be space-like separated from neither of those events.  Furthermore, \textit{no superdeterminism} requires that the cause be in the past of those events, since the present elements must be independent of future interventions.  We formalize these facts with the following theorem:

\begin{quote}
\textbf{Local correlation}: A reproducible perfect correlation between two or more space-like separated regions can only be caused by a correlated
set of localized elements of reality in the intersection of the past light-cones of those regions.
\end{quote}

From \textit{local correlation}, we know that because Alice's and Bob's outcomes are perfectly correlated, those outcomes must be determined by correlated localized elements of reality in the intersection of the past light-cones of the two measurements.  \textit{No superdeterminism} also tells us that those elements of reality are independent of Alice's future decision to measure, say, \textit{z}-spin, and similarly for Bob.

\section{Proving Bell's theorem with many worlds} \label{BellMany}

Perhaps the simplest proof of Bell's theorem is due to Greenberger, Horne, and Zeilinger (1989). Their version uses a particular state of three spins \textit{a, b,} and \textit{c} called the \textit{GHZ state}. The three particles are entangled at the source so that they subsequently obey the following empirically confirmed\footnote{The first experiments that confirmed entanglement correlations were performed by Freedman and Clauser (1972) and then Aspect et al. (1982). Tittel et al. (1998) sent entangled photons in fibre optic cables apart by more than 10km and still found the entanglement correlations, showing that they are not affected by distance. Pan et al. (2000) were the first to confirm the entanglement correlations specific to the GHZ state. These experiments were forced to make certain assumptions which led to potential loopholes that EPR could in principle escape from. However, more recently, a number of loophole-free experiments have been performed (e.g. Giustina et al. (2015); Hensen et al. (2015); and Shalm et al. (2015)). The entanglement correlations of quantum mechanics were again confirmed.} correlations:

\begin{quote}
\textbf{Entanglement correlations}:\\ $XXX=-1$, $YYX = +1$, $XYY = +1$, $YXY = +1$ 
\end{quote}

The correlation $XXX=-1$ means that no matter what outcomes are observed by Alice, Bob, and Charlie after they all measure $X$ ($x$-spin), the product of their eigenvalues will be -1. The correlation $YYX=+1$ means that no matter what outcomes are observed by Alice, Bob, and Charlie after Alice and Bob measure $Y$ and Charlie measures X, the product of their eigenvalues will be +1. And so on for $XYY$ and $YXY$.

Particles \textit{a}, \textit{b}, and \textit{c} are prepared in the GHZ state and then sent to Alice, Bob, and Charlie, respectively, at mutual space-like separation. Alice, Bob, and Charlie then each randomly select the $X$ or $Y$ measurement basis for their measurement.

Let us consider a situation in which the following choices of measurement settings are made. Alice chooses to measure $X$ and obtains an outcome $\lambda_X^A = l \in [1,-1]$. Bob and Charlie also measured $X$ and obtained outcomes $\lambda_X^B = m \in [1,-1]$ and $\lambda_X^C = -\lambda_X^A \lambda_X^B = -lm$ respectively, where the form of Charlie's value is required to satisfy entanglement correlation $XXX=-1$.  Specifically, this means that there exists an Alice, a Bob, and a Charlie who experienced these correlated outcomes while they were at mutual space-like separation, and who may later report these outcomes to one another.

Because of \textit{local correlation}, explaining the perfect entanglement correlation requires that what determines their outcomes, $\lambda_X^A$, $\lambda_X^B$, and $\lambda_X^C$ are \textit{localized elements of reality}, and that the correlation between the three was caused by correlated localized elements of reality at the source where the state was prepared.  To summarize, we can say that based on these choices of measurement, the outcomes must be determined by the following \textit{localized elements of reality}:

\begin{quote}
\textbf{Choice Set 1}:\\
$C1_A: \lambda_X^A = l \in [1,-1]$\\
$C1_B: \lambda_X^B = m \in [1,-1]$\\ 
$C1_C: \lambda_X^C = -\lambda_X^A \lambda_X^B = -lm$ [from $XXX=-1$]
\end{quote}

Now consider the situation in which Alice and Bob instead chose to measure $Y$. Due to \textit{local causality}, their choices cannot change Charlie's localized element, $\lambda_X^C$. Due to \textit{no superdeterminism}, Charlie's localized element cannot depend in advance on Alice and Bob's future choice of measurement settings. Consequently, these alternate choices cannot make any difference at all to the Charlie who experienced outcome $\lambda_X^C = -lm$ in \textit{Choice Set 1}. Alice now gets outcome $\lambda_Y^A = n \in [1,-1]$ and Bob now gets outcome $\lambda_Y^B = \lambda_Y^A \lambda_X^C = -lmn$, where the form of $\lambda_Y^B$ is required to obey entanglement correlation $YYX = +1$. Again, the existence of an Alice, Bob, and Charlie who got these correlated outcomes at space-like separation requires that $\lambda_Y^A$ and $\lambda_Y^B$ are localized elements whose correlation originated at the source. To summarize:

\begin{quote}
\textbf{Choice Set 2}:\\
$C2_A: \lambda_Y^A = n \in [1,-1]$\\
$C2_B: \lambda_Y^B = \lambda_Y^A \lambda_X^C = -lmn$ [from $YYX = +1$]\\ 
$C1_C: \lambda_X^C = -\lambda_X^A \lambda_X^B = -lm$ [from Set 1]
\end{quote}

Now consider the situation in which Alice and Charlie measured $Y$ while Bob measured $X$. It follows that we would still have the Alice who experienced $\lambda_Y^A = n$ and the Bob who experienced $\lambda_X^B = m$, and in order to obey $YXY = +l$, Charlie must now get $\lambda_Y^C = mn$, with the correlation between the three originating at the source.  To summarize:

\begin{quote}
\textbf{Choice Set 3}:\\
$C2_A: \lambda_Y^A = n$ [from Set 2]\\
$C1_B: \lambda_X^B = m$ [from Set 1]\\ 
$C3_C: \lambda_Y^C = mn$ [from $YXY=+1$]
\end{quote}

Finally, consider the situation in which Alice measured $X$, while Bob and Charlie measured $Y$.  It is a fact that there must exist a Bob who experienced $\lambda_Y^B = -lmn$ and a Charlie who experienced $\lambda_Y^C = mn$, so there must exist an Alice who experienced $\lambda_X^A = \lambda_Y^B\lambda_Y^C = -l$ in order to satisfy the entanglement correlation $XYY = +1$. This gives:

\begin{quote}
\textbf{Choice Set 4}:\\
$C1_A: \lambda_X^A = \lambda_Y^B\lambda_Y^C = -l$ [from $XYY=+1$]\\
$C2_B: \lambda_Y^B = -lmn$ [from Set 2]\\ 
$C3_C: \lambda_Y^C = mn$ [from Set 3]
\end{quote}

Here we have labelled Alice's $X$ measurement $C1_A$ because it must be the same measurement that is described in Set 1. But notice that the results contradict. We began with a situation in which Alice experiences outcome $\lambda_X^A = l \in [1,-1]$. By simply respecting the entanglement correlations and imposing \textit{local causality} and \textit{no superdeterminism} we have derived the conclusion that in the exact same situation, Alice experiences outcome $\lambda_X^A = -l$, and thus the \textit{one world} axiom is violated, which proves our reformulated version of Bell's theorem.

To be clear, any \textit{truly local} physical theory must predict that there are two Alices after the measurement, one who experienced outcome $\lambda_X^A = +l$, and another who experienced outcome $\lambda_X^A = -l$. The \textit{localized element of reality} $\lambda_X^A = (l,-l)$ is therefore multivalued, predicting the existence of multiple distinct copies of Alice, who are evidently also invisible to one another.

And since we know there exists an Alice who got $l$ and a Bob who got $-lmn$, there must be a Charlie who got $-mn$ in addition to the Charlie who got $mn$. Repeated applications of these arguments shows that every measurement setting results in (at least) two copies of each observer. We thus have a many worlds ontology.

Importantly, we have arrived at this conclusion without ever invoking a wavefunction in Hilbert space.  Only \textit{local causality}, \textit{no superdeterminism}, and experimental facts were used to prove the existence of many worlds for each macroscopic observer.  We now move to the question of whether a self-consistent \textit{truly local} many worlds theory can actually exist.

\section{Global worlds} \label{Global}

For the purpose of this paper, we define a \textit{global world} as one in which every macroscopic property has a single definite value.

In the following subsections we will discuss two variant many \textit{global worlds} models. In the first, the global worlds are created by branching (\textit{global branching}), in the second, there is no branching -- each global world has always existed (\textit{global divergence}).

\begin{figure}[t]
    \centering
    \includegraphics[width=3in]{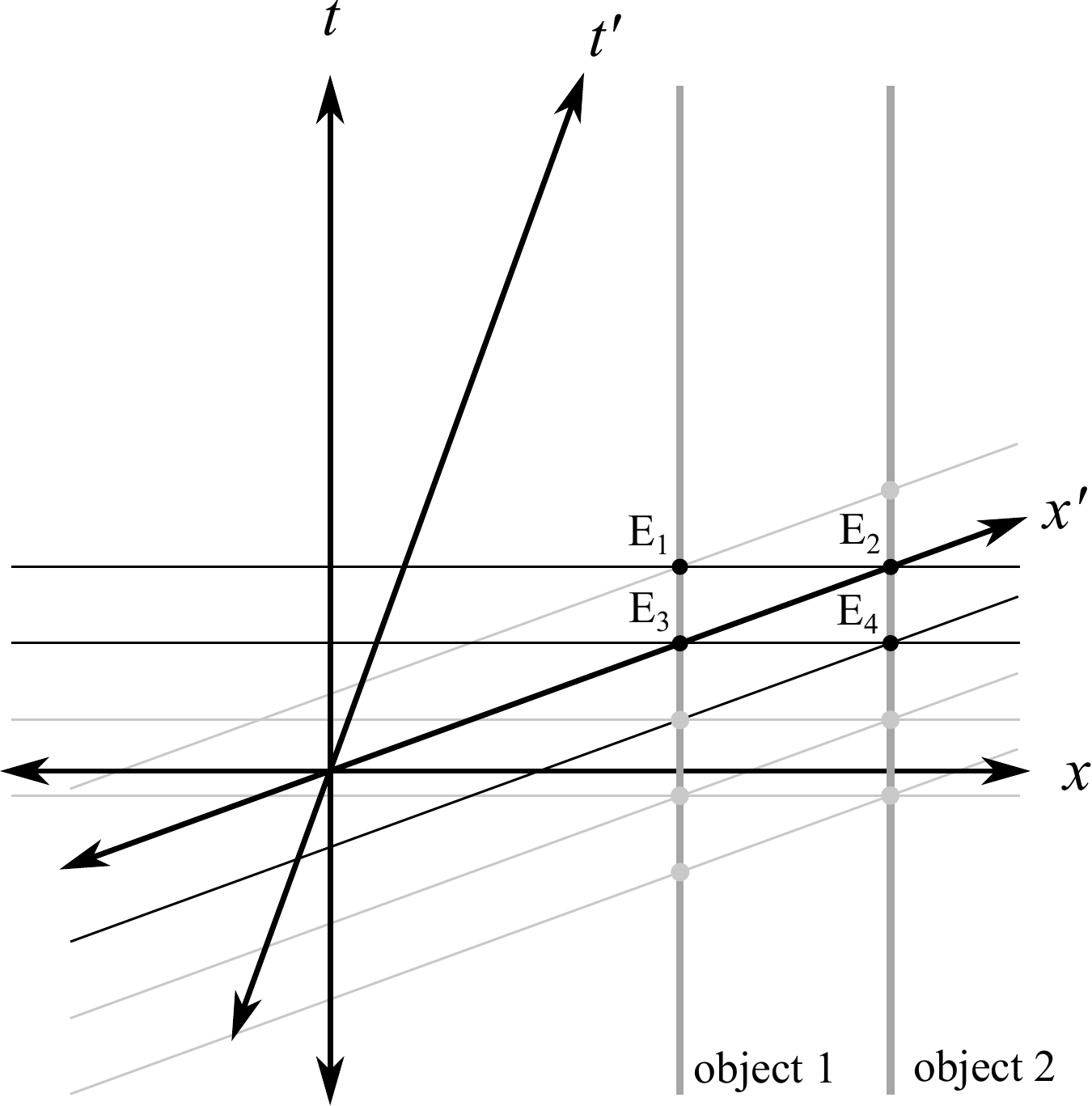}
    \caption{A space-time diagram showing two objects in two different Lorentz frames, $S$ and $S'$.  Consider the case that object 1 is made to branch at event E$_1$.  In frame $S$, this causes object 2 to branch at E$_2$, because E$_1$ and E$_2$ are simultaneous in $S$. But in frame $S'$ E$_2$ is not simultaneous with E$_1$, and thus object 2 does not branch there, which makes the branching situation at event E$_2$ ill-defined. In order to obtain a model with well-defined events in space-time, we must either choose a preferred Lorentz frame in which branching occurs (e.g. only $S$ or only $S'$), or we must adopt the view that if branching occurs in any frame, then the object has branched in all frames.  If we take this latter view, then object 1 branching at E$_1$ causes object 2 to branch at E$_2$ in $S$, which causes object 1 to branch at E$_3$ in $S'$, which causes object 2 branch at E$_4$ in $S$, and so on, until both objects are branched for all time.}
    \label{fig:branching}
\end{figure}

\subsection{Global branching}\label{Global branching}

The distinction between global and local branching was introduced by Sebens and Carroll (2018, pp.33-5). Later, we will argue that even what they call ``local branching", is not really local. We will therefore refer to it as \textit{semi-local branching}. We can understand the distinction with the following example. If Alice has measured the $x$-spin of particle $a$, while Bob is yet to measure particle $b$, then we can represent the overall quantum state as follows:

\begin{equation}\label{local}
\frac{1}{\sqrt 2}(\Ket{\uparrow_x}_{\rm a}\Ket{``\uparrow_x"}_{\rm A}\Ket{\uparrow_x}_{\rm b}+
\Ket{\downarrow_x}_{\rm a}\Ket{``\downarrow_x"}_{\rm A}\Ket{\downarrow_x}_{\rm b}) \otimes \Ket{R}_{\rm B}
\end{equation}

Equivalently, we can represent it like this:

\begin{equation}\label{global}
\frac{1}{\sqrt 2}(\Ket{\uparrow_x}_{\rm a}\Ket{``\uparrow_x"}_{\rm A}\Ket{\uparrow_x}_{\rm b}\Ket{R}_{\rm B}+
\Ket{\downarrow_x}_{\rm a}\Ket{``\downarrow_x"}_{\rm A}\Ket{\downarrow_x}_{\rm b}\Ket{R}_{\rm B})
\end{equation}

According to \textit{semi-local branching}, (\ref{local}) is a more transparent representation of the ontology: since Alice has not interacted with Bob, Bob does not branch. But according to \textit{global branching}, branching happens everywhere if it happens somewhere, so if Alice branches then Bob branches instantaneously. In that case, (\ref{global}) is a more transparent representation of the ontology, since it represents two (qualitatively identical yet numerically distinct) Bobs. We now consider whether the global branching ontology is \textit{truly local}. We will argue that it is not.

In the relativistic context, the idea that branching happens everywhere if it happens somewhere entails that branching occurs simultaneously throughout a space-like hypersurface. One way of making sense of this idea involves postulating a preferred Lorentz frame such that branching occurs simultaneously everywhere in a space-like hypersurface in that frame. This would allow branching at an event to be well-defined (frame-invariant).  If there is no preferred frame, but branching still occurs simultaneously throughout space-like hypersurfaces, then there is a different narrative of branching events in each Lorentz frame -- which renders branching at an event ill-defined (frame-relative).\footnote{This appears to be the option that Sebens and Carroll opt for: ``In a relativistic context, one natural option would be to say that although the number of copies of an observer is frame-relative, the probabilities assigned to measurement outcomes are not." (2018, p35 footnote 7.)} In this case, the ontology includes \textit{all} of the different (well-defined) branching narratives for all Lorentz frames (e.g. somewhere object 2 branches at E$_2$, and somewhere it does not, as in Fig. \ref{fig:branching}.), with objects branching simultaneously everywhere in a space-like hypersurface of a given frame -- independently of when they branch in any other frame.

Now, whether we consider a preferred frame, or just any frame in which global branching occurs, it seems that \textit{local causality} is violated, since Alice causes Bob to branch instantaneously. To contest this, one could claim that Alice is not the cause of Bob's branching, because their branching has a local common cause. However, for this explanation to work, Bob would have to carry with him a local element of reality that determines that he will branch at the right time (i.e. simultaneously with Alice's measurement). Such an element would then enable us to predict when Alice's measurement will cause the composite Alice-Bob system to branch, which violates \textit{no superdeterminism}.  Hence, any global branching model which entails that branching occurs simultaneously everywhere in a space-like hypersurface must violate at least two of the axioms of Bell's theorem, \textit{one world}, and either \textit{local causality} or \textit{no superdeterminism}.

A remaining possbility for making sense of global branching does not posit a preferred frame, yet aims to treat branching as well-defined (frame-invariant). But under these assumptions, `branching anywhere means branching everywhere'  implies `branching at any time means branching at all times'. To see this, consider two always-distant objects in two different frames, as in Fig. \ref{fig:branching}. This would result in a \textit{divergent worlds} model since all branches must have always pre-existed.\footnote{Although they use the language of `branching', Sebens and Carroll (2018, p31) note that they do not mean to prejudge the global branching versus global divergence issue, since their proof of the Born rule can be implemented on either picture.} In the next section, we show that even a model of this type cannot be \textit{truly local}.

\subsection{Global divergence} \label{sec:GlobDiv}

 In a \textit{branching worlds} ontology, a measurement \textit{splits} one world into multiple worlds. But in a \textit{divergent worlds} ontology, a measurement \textit{reveals} pre-existing multiple worlds, which were, up until the measurement, qualitatively identical. 
 
 Divergent worlds entails that all the worlds that will ever need to exist to accommodate the different experiences of macroscopic objects have always existed. Subsets of such worlds are qualitatively identical until corresponding macroscopic objects experience different outcomes, at which point they diverge.\footnote{This notion of \textit{diverging} worlds subsumes the notion of \textit{overlapping} worlds (see Saunders and Wallace (2008) and Saunders (2010)). Although these authors go on to distinguish these notions, the argument of this section applies to both.} But nothing thereby happens at locations remote from these macroscopic objects. Consequently, local elements of reality are not required at every point in space, as in global branching. Divergent worlds thus seem to have a better chance of satisfying the remaining axioms of Bell’s theorem, but things are not so simple. 

Let us reformulate Bell’s theorem for the new
scenario.

\begin{quote}
\textbf{Local Causality}: There can be no cause and effect between space-like separated events.

\textbf{Outcome Definiteness}: Every measurement has a single definite outcome in each global world.

\textbf{No Superdeterminism}: The elements of reality that determine a system's response to interventions do not determine what interventions will occur on that system.
\end{quote}

Here, we have simply replaced the \textit{one world} axiom with the equivalent axiom for many global worlds, which we call \textit{outcome definiteness}. We can label and index the different global worlds in terms of the outcomes that macroscopic objects experience
within them. In each world, the entanglement correlations of any entangled state that was measured must be obeyed. Returning to the GHZ example, if Alice, Bob, and Charlie all measured $X$, then there must be four different types of global worlds, corresponding to outcomes $w_1 \equiv (+1^A +1^B;-1^C)$,
$w_2 \equiv (+1^A;-1^B;+1^C)$, $w_3 \equiv (-1^A;+1^B;+1^C)$, $w_4 \equiv (-1^A;-1^B;-1^C)$, since these are the four combinations that satisfy $XXX = -1$.

Since each global world must obey the $XXX = -1$ correlation, even while the Alice, Bob, and Charlie within that world remain
at space-like separation, the correlation must be caused by correlated local elements of reality at the source. We already know that the \textit{localized elements of reality}, $\lambda_X^A$, $\lambda_X^B$, and $\lambda_X^C$ must be multivalued. But this new requirement means that they must each contain four distinct
values, with each one indicating the outcome experienced in a particular global world. In this example we have the following \textit{localized elements of reality}:

\begin{quote}
$\lambda_X^A = (+1^{w_1}, +1^{w_2}, -1^{w_3}, -1^{w_4})$, 

$\lambda_X^B = (+1^{w_1}, -1^{w_2}, +1^{w_3}, -1^{w_4})$, and

$\lambda_X^C = (-1^{w_1}, +1^{w_2}, +1^{w_3}, -1^{w_4})$.
\end{quote}

Now, from \textit{no superdeterminism} it follows that if, on this same run, Alice and Bob had instead decided to
measure $Y$, while Charlie still measures $X$, then after the measurement there will still exist a Charlie in each of the
four worlds $w_1$, $w_2$, $w_3$, and $w_4$, who experienced outcomes $\lambda_X^C = (-1^{w_1}, +1^{w_2}, +1^{w_3}, -1^{w_4})$, which means that Alice
and Bob must now experience $Y$ outcomes in \textit{the same four} global worlds.

Since the correlation $YYX = +1$ must be obeyed in each of the four \textit{global worlds}, Alice must have a \textit{localized element of reality} $\lambda_Y^A
= (l^{w_1}, m^{w_2}, n^{w_3}, o^{w_4})$, and Bob must have a \textit{localized element of reality} $\lambda_Y^B = (-l^{w_1}, m^{w_2}, n^{w_3}, -o^{w_4})$, with $l$, $m$, $n$, $o \in [+1,-1]$.

Now, if, on the same run, Alice and Charlie had decided to measure $Y$, while Bob had measured $X$, and $YXY = +1$ is obeyed in each of the four worlds, then Charlie must also have a \textit{localized element of reality} $\lambda_Y^C = (l^{w_1}, -m^{w_2}, n^{w_3}, -o^{w_4})$.

Now, finally, if Alice had decided to measure $X$, while Bob and Charlie had measured $Y$, the correlation $XYY = +1$ must be obeyed. But if we consider the known \textit{localized elements of reality}, we see that

\begin{quote}
$\lambda_X^A \lambda_Y^B \lambda_Y^C$ = ($ (+1 \cdot -l \cdot l)^{w_1}$, $(+1 \cdot m \cdot -m)^{w_2}$, $(-1 \cdot n \cdot n)^{w_3}$, $(-1 \cdot -n \cdot -n)^{w_4}$) 

= ($-1^{w-1}$, $-1^{w_2}$, $-1^{w_3}$, $-1^{w_4})$ $\neq +1$.
\end{quote}

\noindent This contradiction shows that no \textit{truly local} global worlds model can reproduce the empirical correlations.

\section{Semi-local worlds}  \label{Semi}

\subsection{Semi-local branching}  \label{sec:SLocBranch}

Recall the distinction between global and semi-local branching, in terms of the situation described by equations ($\ref{local}$) and ($\ref{global}$). We can also consider the situation right after Bob performs an $X$ measurement:

\begin{equation}\label{local2}
\Ket{\psi} = \frac{1}{\sqrt 2}(\Ket{\uparrow_x}_{\rm a}\Ket{``\uparrow_x"}_{\rm A}\Ket{\uparrow_x}_{\rm b}\Ket{``\uparrow_x"}_{\rm B}+
\Ket{\downarrow_x}_{\rm a}\Ket{``\downarrow_x"}_{\rm A}\Ket{\downarrow_x}_{\rm b}\Ket{``\downarrow_x"}_{\rm B})
\end{equation}

Both global and local branching agree that this is the correct expression, but the interpretation is different if signals have not reached Alice from Bob's site (or Bob from Alice's site). Alice has split into two descendants, so has Bob, but there is not yet a world containing an Alice and a Bob. The branching caused by a measurement event spreads through the environment as systems interact with one another, creating a wavefront whose speed is bounded by \textit{c}. A world containing a Bob and an Alice is only created when the wavefront from Alice's measurement meets the wavefront from Bob's measurement. 

Semi-local branching is designed to avoid measurement-induced action at a distance. Consequently, many authors have deemed the theory local (as we discuss in the next section). Here we examine whether semi-local branching models are \textit{truly local}. 

Recall that from \textit{local causality} and \textit{element of reality} we were able to deduce \textit{localized element of reality}. We now apply this principle to the region where Alice and Bob meet to compare their results, which we will treat as the moment at which a world containing both an Alice and a Bob is created, i.e., when the two wavefronts meet.

After her measurement, Alice branches into descendants with different outcomes. Focus on a descendant that obtained spin up. She can predict with certainty that if Bob measured $X$ and she meets him, then he will have found spin up too. There must therefore be a localized element of reality at their meeting event which determines that the up Alice meets an up Bob (and not a down Bob, who is also present in space-time; the down Bob meets a down Alice). This element, $\lambda^{AB}$, enforces the entanglement correlation encoded in ($\ref{local2}$). 

This entanglement correlation originated at the entanglement event of particles $a$ and $b$. For if Alice and Bob had performed their measurements prior to the entanglement, then the correlation would not manifest at their meeting and there would be no reason to posit $\lambda^{AB}$. Moreover, Alice and Bob can perform their measurements arbitrarily soon after the entanglement event, and the correlation would manifest. 

We now consider what was needed to create $\lambda^{AB}$. For this we invoke \textit{local causality} once again, to define a constraint on how localized elements can be created:

\begin{quote} 
\textbf{Localized element synthesis}: A new localized element is created in a region only by existing localized elements in that region.
\end{quote}

If \textit{localized element synthesis} were not true, then a localized element could be created by space-like separated elements of reality. 

The localized element $\lambda^{AB}$ enforces a correlation concerning what outcome Alice reports to Bob (and vice versa). To determine what created $\lambda^{AB}$, we must trace back through Alice and Bob's past, and consider what events could have created $\lambda^{AB}$. 

After Alice measures $a$, she can predict with certainty that if she measures $a$ again, she will get the same result. There must therefore be an element of reality $\lambda^{Aa}$, localized to Alice's region, that determines this response. Similarly, after Bob measures $b$, he can predict with certainty that he will get the same result, so his region contains $\lambda^{Bb}$. However, these elements are insufficient to enforce the correlation at Alice and Bob's meeting, since they are consistent with the absence of $\lambda^{ab}$.

After Alice measures $a$, she can predict with certainty what outcome she would get if she then measured Bob's particle $b$ before Bob gets to it. 
So there must be an element $\lambda^{Ab}$ that determines this response to that intervention. Crucially, $\lambda^{Ab}$ must have been created when Alice measured $a$, and so by \textit{localized element synthesis}, must be located in this region -- for if Alice had measured $b$ before she measured $a$, she would have been unable to predict the result of the $b$ measurement.  

Similarly, after Bob measures $b$, he can predict what he would get if he measured $a$ before Alice gets to it.  So there must be an element $\lambda^{Ba}$ that determines that response to that intervention, and is localized to Bob's region. It is $\lambda^{Aa}$, $\lambda^{Ab}$, $\lambda^{Ba}$, and $\lambda^{Bb}$ that interact at the meeting of Alice and Bob to create $\lambda^{AB}$. 

\begin{figure}
    \centering
    \includegraphics[width = 3in]{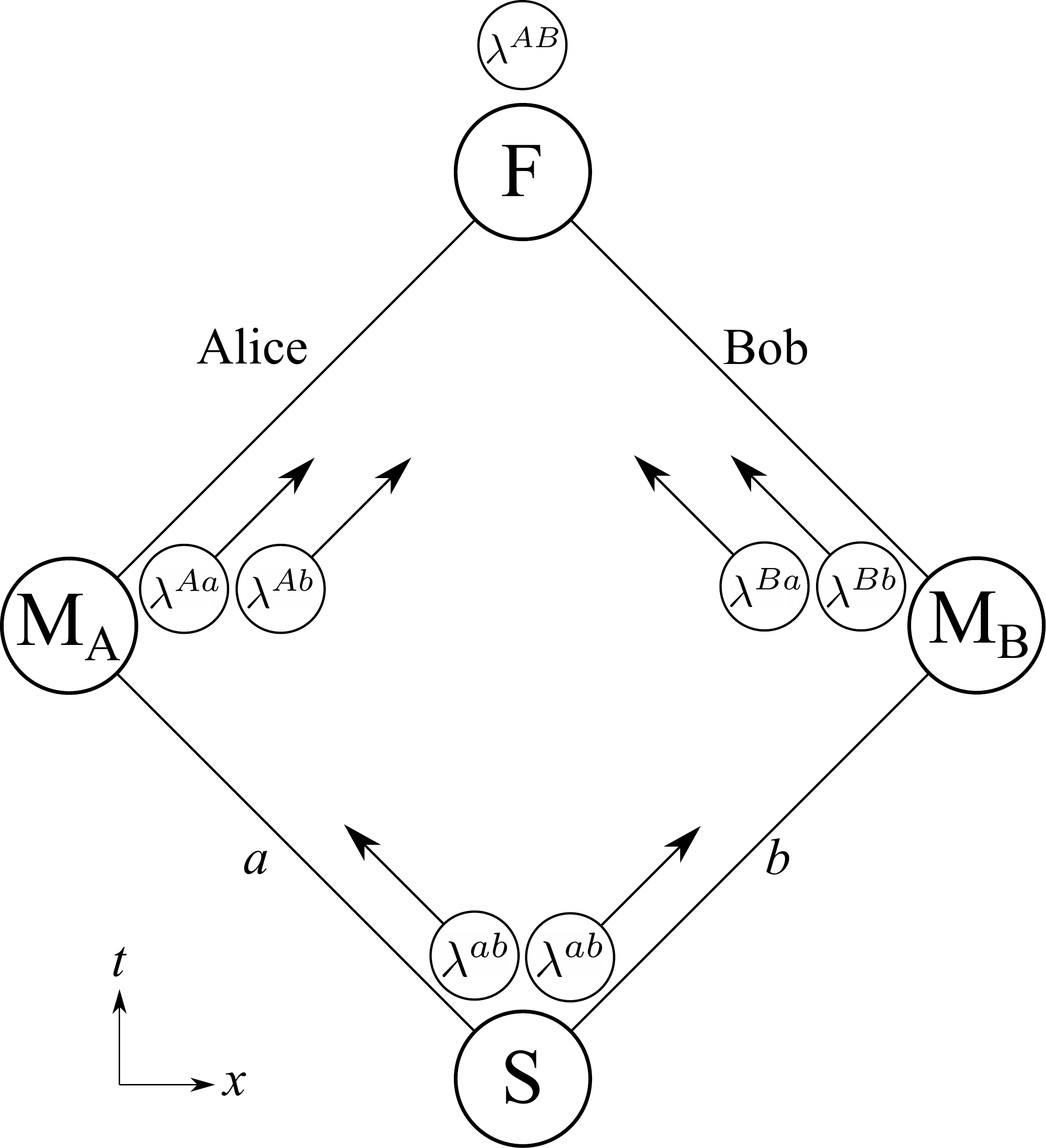}
    \caption{Space-time diagram of the two-spin Bell test, showing the synthesis of localized elements of reality (the $\lambda$s) at the source $S$, the measurements $M_A$ and $M_B$, and the final meeting $F$.  In order to locally synthesize the correlated elements of reality, identical copies of the nonseparable wavefunction $\Psi$ must also be localized elements of reality present at both $M_A$ and $M_B$.  However, the copy at $M_A$ ($M_B$) is instantaneously updated by Bob's (Alice's) measurement, which violates local causality.}
    \label{fig:Synthesis}
\end{figure}

Elements $\lambda^{Aa}$, $\lambda^{Ab}$, $\lambda^{Ba}$, and $\lambda^{Bb}$ in turn depend on $\lambda^{ab}$. We may therefore conclude that all five of these elements were necessary to create $\lambda^{AB}$. The creation process is depicted in Fig. \ref{fig:Synthesis}.

We now ask whether these localized elements, whose existence is required by \textit{true locality}, can be accounted for by semi-local branching. Typically, the ontology of semi-local models is exhaustively described by the wavefunction.\footnote{This statement is neutral as to whether the wavefunction \textit{is} the complete ontology (wavefunction realism) or is the complete \textit{description} of the ontology. In Sec. \ref{Oxford} we apply our argument to another semi-local ontology,  \textit{spacetime state realism}.} However, advocates of semi-local branching models (as well as advocates of global branching models) claim that the wave function enables us to define local descriptions of regions, in the form of reduced density operators.\footnote{See Sebens and Carroll (2018, pp.66-8) and the views discussed in the next section.} Perhaps, then, $\lambda^{Ab}$ could be accounted for by the reduced density operator local to Alice's region and $\lambda^{Ba}$ could be accounted for by the reduced density operator local to Bob's region. But a moment's reflection reveals that this cannot be the case. For the reduced density operator local to Alice's region simply fails to describe the entanglement correlation manifested by Alice's measurement of $b$ after she measures $a$, e.g., Tr$\Big(\big(\rho_{Aa} \Ket{\uparrow}_{\rm Aa}\Bra{\uparrow}_{\rm Aa}\big) \otimes  \big(\rho_{Bb}\Ket{\uparrow}_{\rm Bb}\Bra{\uparrow}_{\rm Bb}\big)\Big) \neq \big|\Bra{\psi}\big(\Ket{\uparrow}_{\rm Aa} \otimes \Ket{\uparrow}_{\rm Bb}\big)\big|^2$, and likewise for any other entangled state.

The entanglement correlations enforced by the localized elements of reality that we have identified, are ultimately contained in the wavefunction, in semi-local branching models. The wavefunction, however, is not a localized entity. Still, if we impose \textit{local causality} and \textit{local synthesis}, then what determines the correlations in these models - $\Psi$ - must be localized to the relevant regions. In that case, there must be multiple identical copies of it, for example, one located at Alice's region, and one located at Bob's region. However, any copy of $\Psi$ must contain all correlation rules, so the copy located in Alice's region contains information about Bob's space-like separated measurement (and vice versa), which violates local causality.

One possible loophole in this argument is that the different localized copies of $\Psi$ are incomplete pieces produced by local common causes with pre-arranged consistency.  However, this is just like assigning the correlated outcomes to global worlds in advance, and thus the argument of Sec. \ref{sec:GlobDiv} already shows that this type of local common cause cannot be consistent with the empirical correlations.

In summary, semi-local models, in which all of nature is described by a single universal wavefunction $\Psi$, cannot be \textit{truly local}.

In recent years, the so-called Oxford Everettians have become the most prominent advocates of semi-local branching. They have also developed an analogous position called \textit{spacetime state realism}. They argue that these are local theories, which contradicts our previous analysis. The next section is therefore devoted to a critical analysis of the Oxford Everettian approach to locality.

\subsection{Oxford Everettianism}\label{Oxford}

Oxford Everettianism refers to a prominent way of expanding upon the many worlds interpretation of Everett [1957], which has been contributed to by many Oxford-affiliated foundations experts. Here we focus on the accounts of locality found in Timpson and Brown (2002), Wallace and Timpson (2010), Wallace (2012 ch.8), and Brown and Timpson (2016).

In Wallace and Timpson (2010) and Wallace (2012 ch.8) a semi-local branching model in spacetime is developed, which is dubbed \textit{spacetime state realism} [SSR]. We begin by analyzing this model as it is presented in Wallace (2012, ch.8), since one of the main purposes of his chapter is ``to determine whether Everettian quantum mechanics is in a relevant sense \textit{nonlocal}" (p292). 

Wallace distinguishes two notions of nonlocality. The first is  \textit{action at a distance} which occurs when ``given two systems $A$ and $B$ which are separated in space, a disturbance to $A$ causes an immediate change in the state of $B$, without any intervening dynamical process connecting $A$ and $B$" (p.293). This seems equivalent to our definition of nonlocality which states that there can be no cause and effect between space-like separated events.\footnote{Wallace rightly distinguishes \textit{action at a distance} from \textit{no signalling}, and states that the latter entails the former but not vice versa.}

Wallace's second notion of nonlocality is \textit{nonseparability}, which occurs when, ``given two regions $A$ and $B$, a complete specification of the states of $A$ and $B$ separately fails to fix the state of the combined system, in addition to the facts about the two individual systems" (p293). The question of nonlocality then breaks down to the questions of whether the theory exhibits action at a distance, and whether it exhibits nonseparability. 
SSR is intended answer these questions. It is a semi-local branching model where the universe can be divided into subsystems. The density operator of each subsystem represents ``the intrinsic properties which that subsystem instantiates, just as the field values assigned to each spacetime point in electromagnetism represented the (electromagnetic) intrinsic properties which that point instantiated." (p299). Regions of spacetime (and their unions) are what has these intrinsic properties.

SSR is then said to be local in the sense of no action at a distance, but nonlocal in the sense of exhibiting nonseparability (pp.302-305). It exhibits nonseparability because knowing the density operators of regions $A$ and $B$ does not suffice to fix the density operator of $A\cup B$. But it avoids \textit{action at a distance} because ``the quantum state of any region depends only on the quantum state of some cross-section of the past light cone of that region. Disturbances cannot propagate into that light cone" (p302). Nonseparability is described as not being a matter of dynamics, but of ontology (p293), and so the ultimate conclusion is that Everettian quantum theory is a theory of local interactions and nonlocal states (p305).

The claim that nonseparability is not a matter of dynamics, but of ontology, is puzzling. We postulate entities in our ontology because of the dynamical effects we think those entities have on ourselves and our measuring instruments. The exception is abstract entities such as numbers; some postulate that numbers exist on the basis of their essential role in formulating physical theories (Platonism). But Wallace and Timpson are not thinking of the subsystem states represented by density operators as abstract entities. They are supposed to be intrinsic physical properties of physical systems. Their nonlocal nature must therefore manifest dynamically somehow.

Following Timpson and Brown (2002), Wallace explains that the nonseparable state of $A\cup B$ is to be understood in terms of an entanglement \textit{relation} between the intrinsic state of $A$ and the intrinsic state of $B$. The key claim is then that nonseparable states \textit{can} be dynamically relevant, but \textit{only} when the distance of the entanglement relation is reduced to zero. Wallace provides an instructive analogy:

\begin{quote}
    ``Somewhat picturesquely, we can think of entanglement between states as a string connecting those states, representing the nonlocal relation between them. We can move either end of the string by local interactions, and we can cause the string to `fray' at either end by entangling the system at each end with adjacent systems. But we cannot access the information content of the string - i.e. we cannot set up dynamical processes whose outcomes are dependent on the nonlocal properties represented by the string - without moving the two ends of the string until they coincide. In this way, nonseparability remains fully consistent with dynamical locality" (p304).
    \end{quote}

This analysis contradicts the conclusion of the previous section. To recap, we can break our argument down into four steps. \textit{First}, \textit{local causality} demands that there be localized elements of reality at Alice's and Bob's regions that determine the relevant entanglement correlations. \textit{Second}, the only local objects in semi-local models are those described by local density operators. \textit{Third}, local density operators are insufficient to describe the localized elements. \textit{Fourth}, since in semi-local models, the relevant entanglement correlations are encoded in the entire wavefunction, the view must place copies of it at the distinct regions, which inevitably violates \textit{local causality}. 

We expect that Wallace et. al. will have no objection to the second and third steps of our argument. For they agree that reduced density operators are insufficient to capture the entanglement correlations. Instead of postulating localized copies of the wavefunction to capture these correlations, they instead postulate entanglement relations (the ``strings"). So they reject step four. And this rejection can be traced back to a rejection of step one, since the idea appears to be that these spatially extended strings can do the job of the localized elements, but without violating \textit{local causality}. 

But this seems incorrect. When Alice entangles with particle \textit{a}, she entangles with particle \textit{b}, thereby physically affecting the string connecting \textit{a} and \textit{b}. But there is no local chain of cause and effect running down the string at strictly subluminal speeds. The entire string is affected instantaneously by Alice. And since part of the string is in Bob's region, \textit{local causality} is violated. Wallace effectively replaces this nonlocal \textit{causation} with a nonlocal \textit{relation} between the regions of Alice and Bob. We think it is merely verbal whether we describe the action at a distance as a relation or not. This semantic choice does not obviate the violation of \textit{local causality}. The nonseparability of the Bell states entails nonlocal action in semi-local world models, such as SSR. Such nonseparability is just one out of many ways of creating a violation of \textit{local causality}. 

Brown and Timpson do not distinguish two notions of nonlocality. Instead, they provide a historical analysis of Bell, who began with our \textit{intuitive} constraint on local causality (Bell, 1964), but who ended with a \textit{mathematical} notion, called \textit{factorisability}, which effectively just rules out nonseparability (Bell, 1990). Brown and Timpson (2016) argue that when we move to the context of non-separable theories (e.g. SSR), we can see that ``Bell's mathematically formulated conditions fail properly to capture his intuitive notions of locality" (p113). First, they note that to derive \textit{factorisability} from \textit{local causality}, one needs a common cause principle:

\begin{quote}
``[I]n order to move from Bell's informal statement of local causality---that the proximate causes should be nearby them, and causal chains leading up to these events should lie on or within their past light-cones---to its mathematical formulation in terms of factorisability [...], something like Reichenbach's principle [of common cause] is appealed to: statistical correlations between events must be explained either by direct causal links  between them, or in terms of a Reichenbachian common-cause in the past."  (p113)
\end{quote}

We agree. Indeed, all throughout the paper we have appealed to \textit{local correlation}. But according to Brown and Timpson, in the context of semi-local models, ``the common cause principle is unnatural and unmotivated" (p113):

\begin{quote}
``This is for a very simple reason: in a nonseparable theory there is a \textit{further} way in which correlations can be explained which Reichenbach's stipulations miss out: correlations between systems (e.g., the fact that certain correlations between measurement outcomes \textit{will} be found to obtain in the future) can be explained directly by irreducible relational properties holding between the systems, relational properties themselves which can be further explained in dynamical terms as arising under local dynamics from a previous non-separable state for the total system" (pp.113-4)
\end{quote}

What Brown and Timpson have here missed, is that the \textit{local correlation} constraint is \textit{derivable} from the \textit{local causality} and \textit{no superdeterminism} constraints. This was proved in Sec. \ref{Bell}, where we saw that a reproducible correlation between two space-like separated regions can only have a past cause in a region that is space-like separated from neither of those two regions. Referring to the relevant nonlocal causal processes as ``irreducible relational properties", again, does not obviate the violation of  \textit{local causality}.

Thus, because of their insistence on a single universal wavefunction (or irreducible entanglement relations), empirically consistent semi-local models, such as SSR, cannot be \textit{truly local}.  
 
We now consider local many worlds models that abandon the universal wavefunction.

\section{Local worlds} \label{LocalWorlds}

Local worlds models aim to obey both \textit{local causality} and \textit{no superdeterminism} unequivocally, and to remain consistent with special relativity. In special relativity, the only fundamental objects consistent with local causality are point-like, and these follow world-lines through space-time. Thus, in these models a \textit{local world} is defined as the experience of a point object as it follows its own world-line through space-time.  The experience of a point object is simply a record of the outcomes of interaction events where its past world-line intersected the world lines of other point objects. A given quantum system (such as an electron) is composed of many such point objects. For a given system, point objects which record the same outcomes are qualitatively identical, and thus they effectively experience the same world, whereas those which record different outcomes diverge into different worlds.  The many world-lines of all systems in the universe occupy the same background space-time. 

The experience of a macroscopic object involves frequent intersections of the world-lines of the many point-objects that comprise it, and their respective records of interaction outcomes. The experience of an observer is not identical to the conscious experience of the observer -- the simple recording of information does not by itself constitute consciousness.  In many cases, we can approximate macroscopic objects like Alice and Bob as point-like objects following world-lines through space-time.

Local worlds models have a much less developed literature, mostly using the name \emph{parallel lives} (Brassard and Raymond-Robichaud 2013, 2017, and 2019; Waegell 2017 and 2018).  There is also some variation within the family of local many worlds models, and here we discuss only one such model.  The premise of local worlds models is similar to that of semi-local worlds, in that branching or divergence of worlds begins at a measurement event, and then spreads out as a wave-front whose speed is bounded by $c$.  However, in local worlds, we give up the universal wavefunction, and instead adopt a relativistic wave-field which has a fixed definite value at each point in space-time.  The wave-field contains the bounded wave-fronts explicitly, and is consistent with all of the local elements of reality required by \textit{true locality} (see Fig. \ref{fig:Synthesis}), resulting in a Lorentz invariant model of quantum physics.  

To understand the ontology, we define a subclass of the \textit{localized elements of reality} by contracting the finite region of space-time in which the intervention and response occur down to a single event:

\begin{quote}
\textbf{Local element of reality}: If an intervention and response happen at a single event, and the response can be predicted with certainty, then there is a point-like element of reality located at that event which determines that response.
\end{quote}

The \textit{local element of reality} must be point-like; if it were not, then the instantaneous response could be affected by space-like separated causes, violating \textit{local causality}.

In local branching models, a measurement induces the world-line of a system to branch into multiple new world lines with different experiences, all occupying the same background space-time. In local divergence models, there are already many identical copies of the world-line of a system, and these experience different outcomes when the system is measured, and thus diverge from one another. In what follows we adopt the divergence picture.

In the Bell experiment, when Alice performs her measurement, her world-lines diverge, and everything is \textit{truly local}.  This is accomplished by providing an explicit dynamical model that produces all of the local elements of reality discussed in Sec. \ref{sec:SLocBranch}.  For this, we will make use of the the \textit{wave-field} $\Ket{\psi(\vec{x},t)}$, which is a wave-function-valued field in space-time, in which influences propagate at or below $c$.  The value of $\Ket{\psi(\vec{x},t)}$ at each event is a fixed record of a wavefunction at that event, and that record is independent of anything outside the past light-cone of that event.  $\Ket{\psi(\vec{x},t)}$ is thus a sort of vector field, with a unique Hilbert space vector at each point in space-time.  This makes $\Ket{\psi(\vec{x},t)}$ fundamentally separable in space-time, even if the Hilbert space vector at a given event is entangled.

In this model, the local element of reality at the source event, $S$, is the wave-field at that point in space-time,
 \begin{equation}
\Ket{\psi}^S_{a,b} =\frac{1}{\sqrt 2}(\Ket{\uparrow_x}_{\rm a}\Ket{\uparrow_x}_{\rm b}+
\Ket{\downarrow_x}_{\rm a}\Ket{\downarrow_x}_{\rm b})
\end{equation}
This wavefunction is a fixed record of the source event, which is now carried by both system $a$ and system $b$.  When Alice performs her measurement on $a$ at event $M_{\rm A}$, her wavefunction is entangled with the wavefunction carried by $a$ to create the new wavefunction, 
\begin{equation}
\Ket{\psi}^{M_{\rm A}}_{a,b, A} =U_{\rm Aa} \frac{1}{\sqrt 2}(\Ket{\uparrow_x}_{\rm a}\Ket{\uparrow_x}_{\rm b}+
\Ket{\downarrow_x}_{\rm a}\Ket{\downarrow_x}_{\rm b})\Ket{R}_{\rm A} = \frac{1}{\sqrt 2}(\Ket{\uparrow_x}_{\rm a}\Ket{\uparrow_x}_{\rm b}\Ket{\uparrow_x}_{\rm A}+
\Ket{\downarrow_x}_{\rm a}\Ket{\downarrow_x}_{\rm b}\Ket{\downarrow_x}_{\rm A}).
\end{equation}
This fixed record of event $M_{\rm A}$ is now carried by both Alice and system $a$.  Likewise, when Bob measures system $b$ at event $M_{\rm B}$, the new wavefunction they obtain is, 
\begin{equation}
\Ket{\psi}^{M_{\rm B}}_{a,b, B} =U_{\rm Bb} \frac{1}{\sqrt 2}(\Ket{\uparrow_x}_{\rm a}\Ket{\uparrow_x}_{\rm b}+
\Ket{\downarrow_x}_{\rm a}\Ket{\downarrow_x}_{\rm b})\Ket{R}_{\rm B} = \frac{1}{\sqrt 2}(\Ket{\uparrow_x}_{\rm a}\Ket{\uparrow_x}_{\rm b}\Ket{\uparrow_x}_{\rm B}+
\Ket{\downarrow_x}_{\rm a}\Ket{\downarrow_x}_{\rm b}\Ket{\downarrow_x}_{\rm B}).
\end{equation}
Note that these local values of the wave-field are consistent with the localized elements of reality from Sec. \ref{sec:SLocBranch} and Fig. \ref{fig:Synthesis}, and that they respect local causality, because $\Ket{\psi}^{M_{\rm A}}_{a,b, A}$ contains no information about Bob, and $\Ket{\psi}^{M_{\rm B}}_{a,b, B}$ contains no information about Alice.

When Alice and Bob meet at event $F$, their respective wavefunctions combine to produce the new wavefunction,
\begin{equation}
\Ket{\psi}^{F}_{a,b, B} =U_{\rm Aa}U_{\rm Bb} \frac{1}{\sqrt 2}(\Ket{\uparrow_x}_{\rm a}\Ket{\uparrow_x}_{\rm b}+
\Ket{\downarrow_x}_{\rm a}\Ket{\downarrow_x}_{\rm b})\Ket{R}_{\rm A}\Ket{R}_{\rm B} \label{PsiF}
\end{equation}
\begin{equation}
= \frac{1}{\sqrt 2}(\Ket{\uparrow_x}_{\rm a}\Ket{\uparrow_x}_{\rm b}\Ket{\uparrow_x}_{\rm A}\Ket{\uparrow_x}_{\rm B}+
\Ket{\downarrow_x}_{\rm a}\Ket{\downarrow_x}_{\rm b}\Ket{\downarrow_x}_{\rm A}\Ket{\downarrow_x}_{\rm B}),\nonumber
\end{equation}
and this determines which copies of Alice can meet which copies of Bob.  In the general case, different unitary operations $U_{\rm Aa}$ and $U_{\rm Bb}$ correspond to arbitrary measurement settings.

In summary, we have shown that \textit{truly local} many worlds models are possible, with a separable wave-field $\Ket{\psi(\vec{x},t)}$ in space-time instead of a universal wavefunction $\Psi$ in configuration space.  Examples like this serve as a guide, but the correct generalizations of the Schr\"{o}dinger and Dirac equations for the wave-field are not yet known.

Finally, we note that many local worlds models have strong similarities to several existing models, which violate \textit{true locality}.  First, the many-interacting-world models of Hall et al. (2010) and Schiff and Poirier (2012), which are related to Bohmian mechanics, also describe trajectories for many individual worlds, but each world is a point in configuration space rather than space-time.  Second, the so-called many-worlds interpretation of Schr\"{o}dinger from Allori et al. (2010), which has point objects in space-time as worlds, but which they conclude is nonlocal.

\section{Discussion}  \label{Discussion}

\subsection{Proofs of the Born rule}\label{Born}

The distinctive features of the various many worlds models are often postulated to explain the origin of the Born rule. Here we consider the general structure of Born rule proofs in the many worlds literature, the reliance of some of those proofs on locality principles, and the sense in which local worlds models circumvent the need for such a proof.

On a branching model, it is difficult to make sense of the probabilistic predictions of quantum mechanics. Pre-measurement, Alice might assert ``there is a 0.7 probability that I will see spin-up (rather than spin-down)". But given branching, it seems that Alice should assign probability 1 to there being an Alice descendant that sees spin-up. Pre-measurement Alice is not uncertain of anything; she knows she will branch into descendants that see contrary results in different worlds, and she knows that neither of her descendants is the ``real" Alice---they both have equal claim to being pre-measurement Alice. It is therefore unclear what the ``probability 0.7" is a probability of. This aspect of the problem is often referred to as the \textit{incoherence problem}. 

The divergence model was proposed by Saunders and Wallace (2008) and Saunders (2010) to introduce uncertainty into Alice's pre-measurement situation. In this model, the descendant Alices did not branch from a single ancestor Alice. Instead, the descendant Alices were always numerically distinct. The measurement merely caused the pre-measurement Alices to diverge i.e. become qualitatively distinct. Now any given pre-measurement Alice has something to be uncertain about: ``am \textit{I} an Alice that is located in a world that will have a spin-up result? 

The \textit{quantitative problem} of why Alice should be certain \textit{to degree 0.7} that she is in a spin-up world is not solved by divergent models alone. Proofs of the Born rule are developed to solve the quantitative problem. The Oxford Everettians have developed a well-known decision-theoretic proof (Wallace 2012, P.II). Here, we set this approach aside as we wish to discuss proofs that rely on locality principles.  

In branching models, many approaches deny that Alice is uncertain about the future, but argue that Alice is uncertain about her location in the multiverse often enough to solve the incoherence problem. This approach is taken by McQueen and Vaidman (2019) in the context of semi-local branching and by Sebens and Carroll (2018) in the context of global branching.

Proofs of the Born rule typically have two major steps. The first proves the Born rule in the context of symmetric superpositions whose amplitudes are evenly distributed (e.g. when spin-up is assigned 0.5 probability). The second step proves the Born rule for unequally weighted superpositions (e.g. when spin-up is assigned 0.7 probability).\footnote{This two-step structure is apparent in Wallace (2012, p173). Wallace's second step relies on the ``branching indifference" principle, which is critically analysed in Dizadji-Bahmani (2015).} It is in this second step that both McQueen and Vaidman, and Sebens and Carroll, appeal to locality-based axioms. 

McQueen and Vaidman (2019, p19; 2020) appeal to \textit{local supervenience}: whatever happens in region $A$ depends on the quantum state of this region and its immediate vicinity. Here, the quantum state of a region is defined by its reduced density operator. Under this assumption, \textit{local supervenience} may well be true. But it does not follow that \textit{local causality} is satisfied, as argued in Sec. 6. Still, the weaker \textit{no-signaling principle} would be sufficient in place of \textit{local supervenience}. Indeed, McQueen and Vaidman (2019, Sec. 4.1.) first present a proof of the Born rule in dynamical collapse theories, where the second step appeals to relativistic no-signalling, before arguing that a structurally similar proof is available for semi-local branching models.

Sebens and Carroll (2016, p16), in their crucial second step, appeal to an epistemic constraint on probability assignments (credences). The \textit{epistemic separability principle} (ESP) states that the credence one should assign to being any one of several observers having identical experiences is independent of the state of the environment. Sebens and Carroll think ESP has intuitive appeal. McQueen and Vaidman (2019, pp.22-3) worry that it is unclear why we should believe ESP in a model that allows global branching. The present analysis allows us to raise the concern more directly: why should one's credences only be constrained by the facts of one's local region in a model that is not \textit{truly local}? Until this question is answered, it seems that Sebens and Carroll's proof does not have a solid foundation. 

Finally, local worlds models provide a straightforward explanation of the Born rule. As discussed, there are many numerically distinct but qualitatively identical copies of Alice before the measurement. Any given Alice is therefore uncertain about the future, since any given Alice is uncertain whether she is an Alice that will see spin-up. It therefore makes sense for her to assign probabilities to future outcomes. Furthermore, when Alice measures her spin, the proportion of her world-lines which experience each outcome is stipulated to match the Born rule predictions. There is therefore no need for any additional proof of the Born rule.\footnote{See Albert (1992, p130) for a similar explanation of the Born rule in the context of a many \textit{minds} interpretation of quantum mechanics.}

\subsection{Other interpretations}   \label{Interps}

The reformulated version of Bell's theorem is a \textit{which-way} theorem, in the sense that any empirically consisitent physical theory must violate at least one axiom of the theorem. This makes it a useful tool for categorizing the various interpretations of quantum mechanics, and we attempt this here for a number of well-known interpretations. It seems inevitable that some proponents of these models will disagree with our assessments, but we have endeavored to be as fair and open-minded as possible in the following discussion.

To begin with a recapitulation of this paper, we have proved that all interpretations of quantum mechanics which include a universal wavefunction violate \textit{local causality}.  This includes various Copenhagen-ish interpretations, Bohmian mechanics, and dynamical collapse models, all of which obey the \textit{no superdeterminism} and \textit{one world} axioms.

The consistent (or decoherent) histories interpretation of Griffiths (2003) also seems to follow a universal wavefunction, meaning it violates \textit{local causality}, but the details are hard to pin down.  According to Gell-Mann and Hartle (2012) it obeys the \textit{one world} axiom, even if this world cannot be directly observed, but because the model refuses to consider counterfactual questions, it also seems to deny \textit{no superdeterminism}.

All of the many worlds models we have discussed (except local worlds) also violate \textit{local causality} because of the universal wavefunction, and they all violate the \textit{one world} axiom by definition.  Most many worlds models obey \textit{no superdeterminsm}. 
 
Retrocausal interpretations like those of Price and Wharton (2015) and Aharonov,  Bergmann, and Lebowitz (1964) violate \textit{no superdeterminism}, since elements of reality exist only for observables that will be measured in the future, and not for all observables.  These models both obey the \textit{one world} axiom, and Price and Wharton obey \textit{local causality} while Aharanov et al. seem willing to give it up.
 
 The case of QBism from Fuchs, Mermin, and Schack (2014) is a little harder to pin down, but like other Copenhagen-ish models, it seems that most of its proponents think it should obey the \textit{no superdeterminism} and \textit{one world} axioms, meaning that it must violate \textit{local causality}. On the other hand, QBists do assert that their theory is local. As they say, ``QBist quantum mechanics is local because its entire purpose is to enable any single agent to organize her own degrees of belief about the contents of her own personal experience. No agent can move faster than light: the space-time trajectory of any agent is necessarily timelike. Her personal experience takes place along that trajectory." However,  the question of locality is not a question of personal experience, it is a question of physical ontology. But a physical ontology is something that QBists have yet to provide.  
 
Interestingly, the local worlds model can almost be seen as a version of QBism that provides a physical ontology, wherein \textit{one world} is violated instead of \textit{local causality}; for local worlds models agree that each individual agent moves along their own world-line -- it's just that agents are treated no differently than any other physical system. Furthermore, local worlds models agree that the universal wavefunction does not compose the ontology, but that each agent may have their own personal wavefunction, which they use to predict their future experiences. 

 Local worlds is also similar to Rovelli's (1996) relational interpretation of quantum mechanics, which also denies a single universal wavefunction.  Relational quantum mechanics is also intended to obey \textit{local causality} and \textit{no superdeterminism}, in which case our proof shows that it must be a local many worlds model. Their key disagreement on this point seems to stem from their apparent denial that space-like separation is an objective fact about two objects, but without this, they would appear to have no coherent definition of locality (Smerlak and Rovelli 2007, Martin-Dussaud et. al. 2019, cf. Pienaar 2018).

\section{Conclusions}  \label{sec:Conc}
The main objective of this paper was to generalize and formalize both the explicit and implicit axioms of EPR (1935) and Bell (1964) in the pursuit of a truly local quantum theory. From there, we examined a number of existing many worlds models. We found that none of them are \textit{truly local}, with the exception of local many worlds models.

The fact that $N$-particle quantum theory is defined in $3N$-dimensional configuration space, and not in 3-space, was seen as deeply troubling by Schr\"{o}dinger, Einstein, Lorentz, and many others (Norsen 2017, pp. 118-120), partly because it seemed to make the theory nonlocal.  Here we have proved that a universal wavefunction in configuration space cannot belong to the ontology of any \textit{truly local} model, which validates their intuition.
  
Local many worlds is also the first quantum theory (that we know of) that is \textit{truly local}, and where all the physics is Lorentz invariant and exists in space-time (note that even relativistic field theories handle entanglement using a universal wavefunction in configuration space).

Of course, as we have discussed, there are a number of other self-consistent interpretations of quantum mechanics which are not \textit{truly local} and keep the universal wavefunction as part of their ontology. It therefore remains possible that the wavefunction will prove to be a necessary part of the correct quantum theory.  The purpose of this paper is not to advocate for one interpretation over another, but to establish a clear framework in which any interpretation can be impartially analyzed, and to apply that analysis to several models in order to elucidate their inner workings and highlight their differences.\newline

\textbf{Acknowledgments}:---  This research was supported (in part) by the Fetzer-Franklin Fund of the John E. Fetzer Memorial Trust.

\section{References}
\indent

Aharonov, Y., P.G. Bergmann, and J.L. Lebowitz. 1964. Time symmetry in the quantum process of measurement. \textit{Physical Review}, 134(6B).\\

Albert, D. Z. 1992. \textit{Quantun Mecanics and Experience}. Harvard University Press. \\

Allori, V., S. Goldstein, R. Tumulka, N. Zanghì. 2010. Many worlds and Schrödinger’s first quantum theory. \textit{British Journal for the Philosophy of Science}, 62(1), pp.1-27.\\

Aspect, A., P. Grangier, and G. Roger. 1981. Experimental Tests of Realistic Local Theories via Bell's Theorem. \textit{Phys. Rev. Lett}. 47 (7): 460–3.\\

Bell, J.S. 1964. On the Einstein-Podolsky-Rosen paradox. \textit{Physics} 1:195-200.\\

Bell, J.S. 1987. \textit{Speakable and Unspeakable in Quantum Mechanics}. Cambridge University Press. 2nd Edition 2004.\\

Bell, J.S. 1995. `La Nouvelle Cuisine', in \textit{Quantum Mechanics, High Energy Physics And Accelerators: Selected Papers Of John S Bell (With Commentary)} (pp. 910-928).\\

Bohm, D. 1952. A suggested interpretation of the quantum theory in terms of ``hidden" variables, I and II. \textit{Physical Review} 85, 166.\\

Brassard, G. and Raymond-Robichaud, P., 2013. Can free will emerge from determinism in quantum theory?. \textit{In Is Science Compatible with Free Will? (pp. 41-61). Springer}, New York, NY.\\

Brassard, G. and Raymond-Robichaud, P., 2017. The equivalence of local-realistic and no-signalling theories. \textit{arXiv preprint} arXiv:1710.01380.\\

Brassard, G. and P. Raymond-Robichaud. 2019. Parallel lives: A local-realistic interpretation of ``nonlocal'' boxes. \textit{Entropy}, 21(1), p.87.\\

Brown, H. and C. Timpson. 2016. `Bell on Bell's Theorem: The Changing Face of Nonlocality'. In M. Bell \& S. Gao (Eds.), Quantum Nonlocality and Reality: 50 Years of Bell's Theorem (pp. 91-123). Cambridge: Cambridge University Press. doi:10.1017/CBO9781316219393.008\\

Dizadji-Bahmani, F. 2015. The Probability Problem in Everettian Quantum Mechanics Persists.
\textit{British Journal for the Philosophy of Science} 66 (2):257-283.
\\

Einstein, A., B. Podolsky, and N. Rosen. 1935. Can the quantum-mechanical description of physical reality be considered complete? \textit{Physical Review} 47: 777-780.\\

Everett, H. 1957. Relative state formulation of quantum mechanics. \textit{Reviews of Modern Physics} 29: 454–462. \\

Freedman, S.J. and J.F. Clauser. 1972. Experimental test of local hidden-variable theories. \textit{Physical Review Letters} 28 (938): 938–941.\\

Fuchs, C.A., N. D. Mermin, and R. Schack.  2014 An introduction to QBism with an application to the locality of quantum mechanics. \textit{American Journal of Physics}.\\

Gell-Mann, M. and Hartle, J.B., 2012. Decoherent histories quantum mechanics with one real fine-grained history. Physical Review A, 85(6), p.062120.\\

Ghirardi, G.C., A. Rimini, and T. Weber. 1986. Unified dynamics for microscopic and macroscopic systems. \textit{Physical Review} D 34, 470.\\

Giustina, M. et al. 2015. Significant-loophole-free test of Bell’s theorem with entangled photons. \textit{Physical Review Letters} 115.25: 250401.\\

Greenberger, D.M., M.A. Horne, and A. Zeilinger. 2007. \textit{Going beyond Bell's Theorem.} arXiv:0712.0921.\\

Griffiths, R. 2003. \textit{Consistent Quantum Theory}. Cambridge University Press.\\

Hall, M.J., D.A. Deckert, H.M. Wiseman. 2014. Quantum phenomena modeled by interactions between many classical worlds. \textit{Physical Review X}, 4(4), p.041013.\\

Hensen, B, et al. 2015. Loophole-free Bell inequality violation using electron spins separated by 1.3 kilometres. \textit{Nature} 526.7575: 682.\\

Martin-Dussaud, P., C. Rovelli, and F. Zalamea. 2019. The Notion of Locality in Relational Quantum Mechanics. \textit{Found Phys} 49: 96–106.\\

McQueen, K.J. and L. Vaidman. 2019.
In defence of the self-location uncertainty account of probability in the many-worlds interpretation.
\textit{Studies in History and Philosophy of Modern Physics}, 66:14-23.\\

McQueen, K.J. and L. Vaidman. 2020. `How the Many Worlds Interpretation brings Common Sense to Paradoxical Quantum Experiments', in \textit{Scientific Challenges to Common Sense Philosophy}, R. Peels, J. de Ridder R. van Woudenberg (eds). London Routledge.\\ 

Myrvold, W., M. Genovese, A. Shimony. 2019. Bell's Theorem. \textit{The Stanford Encyclopedia of Philosophy (Spring 2019 Edition)}, Edward N. Zalta (ed.),\\ URL = {https://plato.stanford.edu/archives/spr2019/entries/bell-theorem/}.\\

Norsen, T. 2017. \textit{Foundations of Quantum Mechanics, An Exploration of the Physical Meaning of Quantum Theory}. Springer.\\

Pan, J., D. Bouwmeester, M. Daniell, H. Weinfurter, A. Zeilinger. 2000. Experimental test of quantum nonlocality in three-photon GHZ entanglement. \textit{Nature}. 403 (6769): 515–519.\\

Pearle, P. 1976. Reduction of the state vector by a nonlinear Schr\"{o}dinger equation. \textit{Physical Review} D 13, 857.\\

Pienaar, J. 2018. Comment on `The Notion of Locality in Relational Quantum Mechanics', unpublished manuscript. [arXiv:1807.06457 [quant-ph]]\\

Price, H. and K. Wharton. 2015. Disentangling the quantum world. \textit{Entropy}. 17(11), pp.7752-7767.\\

Rovelli, C. 1996. Relational quantum mechanics. \textit{International Journal of Theoretical Physics}. 35(8), pp.1637-1678.\\

Saunders, S. and D. Wallace. 2008. Branching and uncertainty. \textit{British Journal for the
Philosophy of Science} 59, 293-305.\\

Saunders, S. 2010. Chance in the Everett Interpretation, in \textit{Many Worlds?}, Saunders, S., Barrett, J., Kent, A., \& Wallace, D. (eds.) Oxford University Press.\\

Schiff, J. and B. Poirier. 2012. Communication: quantum mechanics without wavefunctions. \textit{The Journal of chemical physics}, 136(3), p.031102.\\

Sebens, C.T. and Carroll, S.M. 2018. Self-locating Uncertainty and the Origin of Probability in Everettian Quantum Mechanics. \textit{The British Journal for the Philosophy of Science}, 69, 1: 25–74.\\

Shalm, L.K. et al. 2015. Strong Loophole-Free Test of Local Realism. \textit{Physical Review Letters} 115.25: 250402.\\

Smerlak, M. and Rovelli, C., 2007. Relational EPR. \textit{Foundations of Physics}, 37(3), pp.427-445.\\

Tittel, W., J. Brendel, B. Gisin, T. Herzog, H. Zbinden, N. Gisin. 1998. Experimental demonstration of quantum-correlations over more than 10 kilometers. \textit{Physical Review} A 57 (5): 3229–3232.\\

Timpson, C.G. and H. Brown. 2002. 
'Entanglement and Relativity' in \textit{Understanding Physical Knowledge.} V. Fano and R. Lupacchini (eds.), University of Bologna, CLUEB; quant-ph/0212140.\\

Waegell, M., 2017. Locally causal and deterministic interpretations of quantum mechanics: parallel lives and cosmic inflation. \textit{Quantum Studies: Mathematics and Foundations}, 4(4), pp.323-337.\\

Waegell, M., 2018. An Ontology of Nature with Local Causality, Parallel Lives, and Many Relative Worlds. \textit{Foundations of Physics}, 48(12), pp.1698-1730.\\

Wallace, D. and C.G. Timpson. 2010. Quantum Mechanics on Spacetime I: Spacetime State Realism. \textit{The British Journal for the Philosophy of Science}, 61(4): 697-727.\\

Wallace, D. 2012. \textit{The Emergent Multiverse: Quantum Theory according to the Everett Interpretation.} Oxford: Oxford University Press.\\

\end{document}